\documentclass[runningheads]{llncs}
\usepackage[T1]{fontenc}
\usepackage{graphicx}
\usepackage{listings}
\begin{document}
\title{A Guide to Re-Implementing Agent-based Models: Experiences from the HUMAT Model}
\titlerunning{A Guide to Re-Implementing Agent-based Models}
\author{\"{O}nder G\"{u}rcan\orcidID{0000-0001-6982-5658} \and
Timo Szczepanska\orcidID{0000-0003-2442-8223} \and
Patrycja Antosz\orcidID{0000-0001-6330-1597}}
\authorrunning{\"{O}. G\"{u}rcan et al.}
\institute{
   Center for Modeling Social Systems (CMSS),\\
   NORCE Norwegian Research Center AS,\\
   Universitetsveien 19, Kristiansand, Norway\\
   \email{\{ongu,timo,paan\}@norceresearch.no}\\
   \url{https://www.norceresearch.no/en/research-group/cmss}
}
\maketitle              % typeset the header of the contribution
\begin{abstract}
Replicating existing agent-based models poses significant challenges, particularly for those new to the field. 
This article presents an all-encompassing guide to re-implementing agent-based models, encompassing vital concepts such as comprehending the original model, utilizing agent-based modeling frameworks, simulation design, model validation, and more. 
By embracing the proposed guide, researchers and practitioners can gain a profound understanding of the entire re-implementation process, resulting in heightened accuracy and reliability of simulations for complex systems. 
Furthermore, this article showcases the re-implementation of the HUMAT socio-cognitive architecture, with a specific focus on designing a versatile, language-independent model. 
The encountered challenges and pitfalls in the re-implementation process are thoroughly discussed, empowering readers with practical insights. Embrace this guide to expedite model development while ensuring robust and precise simulations.

\keywords{Agent-based Models \and Replication \and Re-implementation \and Simulation design \and Model calibration \and Model validation \and Best practices}
\end{abstract}
\section{Introduction}

Recognizing the need to build higher quality social simulation tools, the scientific community has made numerous efforts to develop procedures that improve description \cite{grimm_odd_2020}, reusability \cite{tang_code_2020}, rigor and transperency \cite{achter_rat-rs_2022}, and increase confidence in agent-based model (ABM) outputs. 
One essential procedure that deserves more attention as an external model validation method is model replication - re-implementing an existing model based on a representation provided by model builders (e.g., in the form of a natural language conceptual description or source code). 
Even though agent-based modelers have early on recognized replication as "one of the hallmarks of cumulative science" \cite{axelrod_advancing_1997} and was proposed, alongside verification and validation, as an independent test of a model's reliability \cite{zhong_using_2010}, it is most often brought to attention in negative instances of a failure to reproduce results of the original model (e.g.,\cite{will_replication_2008}). 

Since ABMs provide explicit causal explanations of investigated phenomena \cite{antosz_sensemaking_2022}, replication is vital in validating the model's causal consistency. After all, a causal mechanism represented in the ABM is expected to produce the same effects regardless of the language/software of the implementation. However, if it fails to do so, jumping to conclusions about a widespread replication crisis in social simulation (similar to the one in psychology \cite{maxwell_is_2015} might be premature, given how much we still have to learn about the specificity of agent-based modeling as a scientific method. Re-implementing a conceptually identical model in a novel platform can help validate the causal mechanisms explaining the model outcomes and identify software-implicit assumptions that are not an explicit part of the conceptual causal explanation but influence the model outcomes (e.g., \cite{edmonds_replication_2003}). 

Re-implementing existing ABMs in another programming language is a crucial task for researchers and practitioners seeking to enhance the flexibility and scalability of their simulations. 
Until now, various studies have emphasized the importance of re-implementing agent-based models in different programming languages \cite{railsback_concepts_2001,edmonds_replication_2003,an_agent-based_2009,thiele_facilitating_2014,liang_testing_2015}. 
Railsback \cite{railsback_concepts_2001} emphasizes the need for re-implementing models in diverse programming languages to capture and represent the complexity of real-world systems.
Edmond and Hales \cite{edmonds_replication_2003} state that replication can reveal surprising flaws, even in the simplest of models. 
Chattoe-Brown et al. \cite{chattoe-brown_reproduction_2021} emphasize that ensuring such replication becomes even more critical when the model outcomes have the potential to impact individuals' lives significantly. 

Unfortunately, replication of ABMs is underused in practice. 
Zhong and Kim \cite{zhong_using_2010} elaborate on possible challenges that explain why re-implementation is still rare. 
They emphasize that replication is a highly resource-demanding activity with relatively low payoffs in the form of publishable articles, sometimes seen as a trivial activity given to students who take their first steps in coding. 

This article attempts to aid in building procedures that support replication \cite{sansores_agent-based_2005,wilensky_making_2007,zhang_replication_2021}, recognizing the importance of the original research process that starts with the conceptual model. The aim is to report on a systematic process of model replication, sharing good practices and lessons learned from re-implementing the HUMAT socio-cognitive architecture in Python (following an original implementation in NetLogo). 
The following section introduces a systematic guide for re-implementing agent-based models - a step-by-step process of model re-implementation. 
We developed this guide alongside the re-implementing HUMAT in Python case study. Effort was taken to generalize the re-implementation process. The guide proposed here serves as a starting point, aimed to be further developed. 
The article concludes with a short discussion.

\section{Guide for Re-implementing Agent-Based Models}
\label{sec:Re-implementing-Agent-Based-Models}

% (Önder) How models can be re-implemented - provide a guide with steps and what should be paid attention to in each step.

\begin{figure}
\centering
\includegraphics[width=0.99\textwidth]{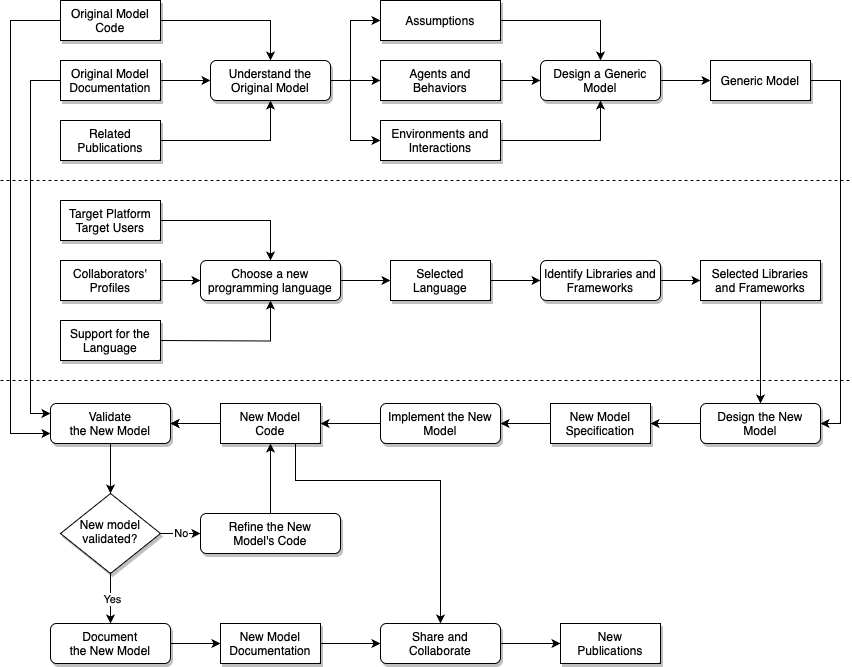}
\caption{The process for re-implementing ABM Models}
\label{fig:ABM-Reimplementation-Process}
\end{figure}

Re-implementing an existing agent-based model in a different programming language involves a series of steps to ensure the new implementation is accurate, efficient, and maintainable. We propose the following systematic approach to guide the re-implementation process (Figure \ref{fig:ABM-Reimplementation-Process}), summarized in the most important steps below.

%For re-implementing an existing agent-based model in another programming language, we propose the following steps (Figure \ref{fig:ABM-Reimplementation-Process}):

\textbf{Understand the original model}: 
Before beginning any re-implementation, it is essential to clearly understand the existing model's functionality and designs \cite{pressman_software_2014}.
This allows the developer to identify potential issues or limitations that should be addressed in the new implementation.
Hence, we need to start by studying the original model's documentation, code, and any related publications and gain a thorough understanding of its objectives, assumptions, agents, behaviors, interactions, environment, and other relevant aspects.

\textbf{Design a generic model}: 
If the original model's documentation is tightly coupled with the original programming language, we must outline a generic model independent of a programming language and framework. The generic model should describe the objective, assumptions, agents, behaviors, interactions, and environment. 
In that sense, applying UML and object-oriented patterns \cite{larman_applying_2004} and pattern-oriented agent modeling \cite{grimm_pattern-oriented_2012} are good candidates.

\textbf{Choose a new programming language\footnote{Note that the initiation of this step is independent from initiation of the other steps and can start at any time.}}:
Choosing the correct programming language can significantly impact its success and depends on several factors \cite{railsback_agent-based_2019,north_agent_2009}. 
%Select a well-suited programming language for the intended agent-based modeling project \cite{railsback_agent-based_2019}. 
The criteria to be considered are the target platform, target users, (if any) partners' experience/preference, and the language's community, library, and support strength. Common choices include Python, Java, and NetLogo \cite{abar_agent_2017}.

\textbf{Identify appropriate libraries or frameworks}: Research and choose libraries or frameworks that are compatible with your chosen programming language and can facilitate agent-based modeling. For example, Mesa for Python \cite{masad_mesa_2015}, Repast for Java \cite{collier_repast_2003}, or NetLogo's built-in constructs/extensions.

\textbf{Design the new model}: Based on the generic model and considering the chosen language and framework, design a new model representing agents, environments, interactions, and behaviors. Consider whether any modifications, adaption of the data structures, or optimizations should be made to the generic model based on the new programming language's capabilities.

\textbf{Implement the new model}: Translate the design model into the chosen programming language, adapting the structure and syntax as needed. Use the chosen libraries or frameworks to help streamline the process.

\textbf{Validate the new model}: Test the new model against the original to ensure it produces the same or similar results \cite{gurcan_generic_2014,gurcan_towards_2011}. 
This may involve comparing outputs, such as agent behaviors, interactions, aggregate patterns, and any performance metrics. Address any discrepancies or issues that arise.

%Optimize the new model: If necessary, refine the new model's code and architecture to improve its efficiency, readability, or maintainability. This might include leveraging language-specific features, optimizing data structures, or implementing parallelism.

\textbf{Document the new model}: Create thorough documentation for the new model, including explanations of its purpose, assumptions, agents, behaviors, interactions, and environment. 
In that sense, the ODD protocol \cite{grimm_odd_2020} or UML-based specifications \cite{larman_applying_2004} can be used.
Include information on any changes or optimizations made during the re-implementation process.

\textbf{Share and collaborate}: Share the new model with the original model's authors and the broader research community through platforms like CoMSES\footnote{CoMSES Model Library, \url{https://www.comses.net/codebases/}, last access on 11/05/2023.}, GitLab, GitHub, and through scientific journals and conferences. Solicit feedback, collaborate on improvements, and contribute to the growing body of knowledge in agent-based modeling.

%Maintain the new model: Continually update and maintain the new model as needed, addressing bugs, incorporating new research findings, or adapting to changes in the field or programming language.

\section{Case Study: Re-Implementing HUMAT}

We have chosen a realistic case study to validate the effectiveness of the proposed re-implementation process.
In the following, we present how we followed the abovementioned guideline (Section \ref{sec:Re-implementing-Agent-Based-Models}) in three subsections.

\subsection{Choosing the Programming Language and Identifying the Libraries/Frameworks}

In our case, the need for re-implementation was driven by the goal of the URBANE project\footnote{URBANE, \url{https://www.urbane-horizoneurope.eu}, last access on 10/05/2023.} that requires combining the elements of two different simulation models: HUMAT \cite{antosz_simulation_2019} (implemented in NetLogo) and MASS-GT \cite{de_bok_empirical_2018} (implemented in Python) into a single simulation model. 
Since the target of the resulting model will be used by our partner who knows Python and integrating HUMAT will be easier if we have a Python version, we decided to re-implement HUMAT in Python.

NetLogo is a well-documented ABM platform that uses a primary object-oriented language with primitives (predefined keywords) to control agents. Python is a general-purpose, high-level programming language. For the URBANE implementation, we used the Mesa ABM framework \cite{masad_mesa_2015}. Mesa extends Python with several functionalities to make programming ABMs more manageable. 
While it is less comprehensive and well-documented than NetLogo, it offers modelers the benefit of accessing many Python libraries.

\subsection{Understanding HUMAT and Designing its Generic Model}

\begin{figure}
\centering
\includegraphics[width=0.99\textwidth]{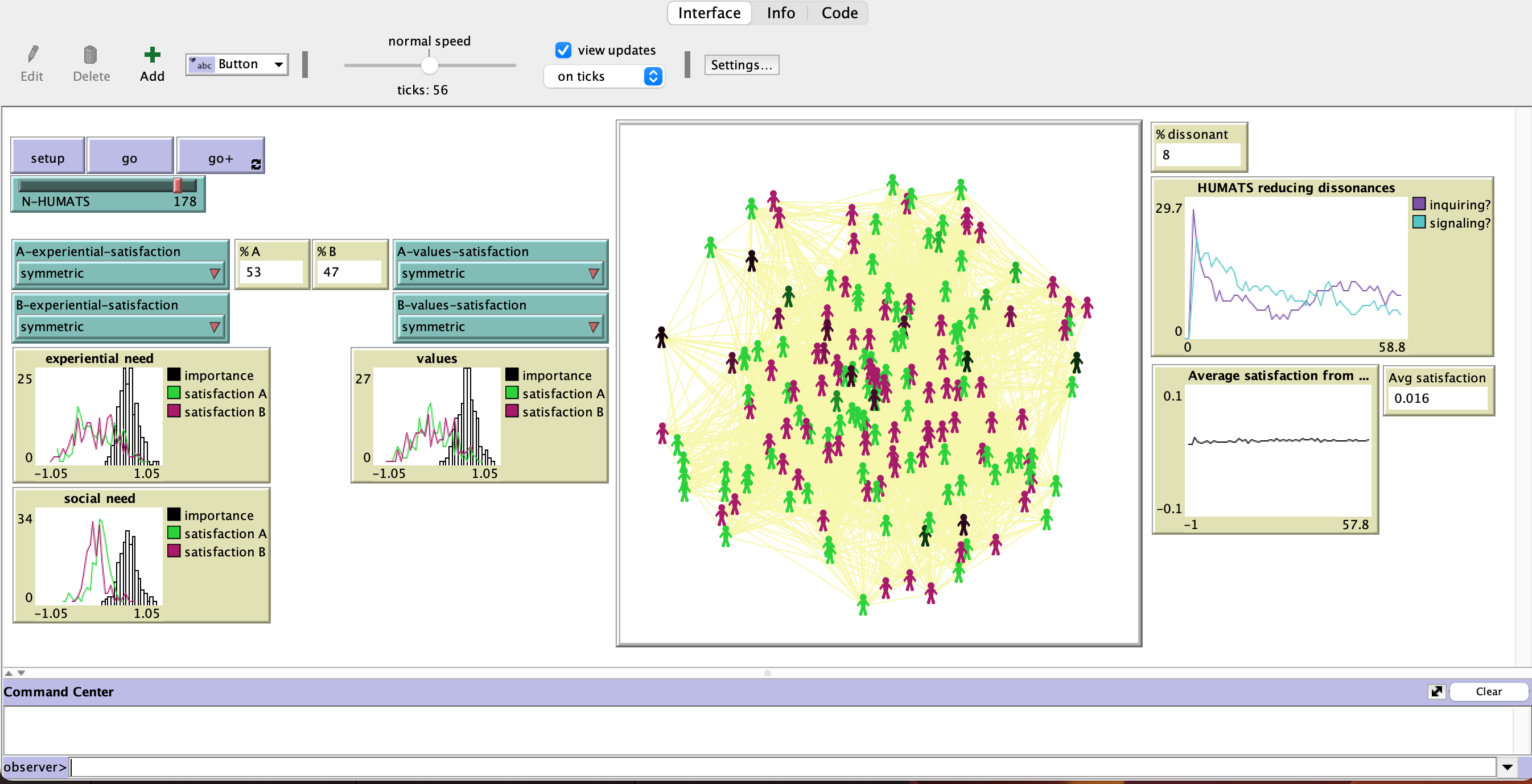}
\caption{The NetLogo version of HUMAT.}
\label{fig:HUMAT-NetLogo-GUI}
\end{figure}

To understand HUMAT, we used the available documents and publications \cite{antosz_simulation_2019,antosz_smartees_2021,czupryna_documenting_2022,squazzoni_hum-e_nodate}, and its corresponding NetLogo version (Figure \ref{fig:HUMAT-NetLogo-GUI}).

As a result of the understanding process, the purpose of the HUMAT model is to represent agents' socio-cognitive process of attitude formation. 
The subjects of the attitude – the options an agent decides between (alternative A and alternative B) are decided by the modeler to fit the research problem that the agent-based model investigates. 

The model is composed mainly of HUMAT agents connected through one or several social networks (i.e., ego networks).
Each HUMAT agent is characterized by a set of needs/motives that are important for the subject of the attitude that can belong to one of three groups: experiential needs, social needs, and values. 
HUMAT agents vary regarding the importance of each motive and how the choice alternatives satisfy each motive. 
When HUMAT agents form their attitude toward a choice alternative, they reflect on how satisfying that alternative is. 
If the alternative satisfies one motive and dissatisfies another motive (i.e., has pros and cons), a HUMAT agent experiences an unpleasant state of dissonance. 
Consequently, that agent faces a dilemma and employs one of two dissonance resolution strategies to maintain cognitive consistency.
Suppose the dilemma is non-social (i.e., the social need to be surrounded by enough like-minded HUMATS is satisfied). In that case, the HUMAT inquires - strives to change its own beliefs by asking the most persuasive alter in the ego network for advice. 
If the dilemma is social (i.e., the social need is dissatisfied), the HUMAT signals to the most gullible alter, striving to persuade them to change their mind. 

\begin{figure}[ht]
\centering
\includegraphics[width=0.80\textwidth]{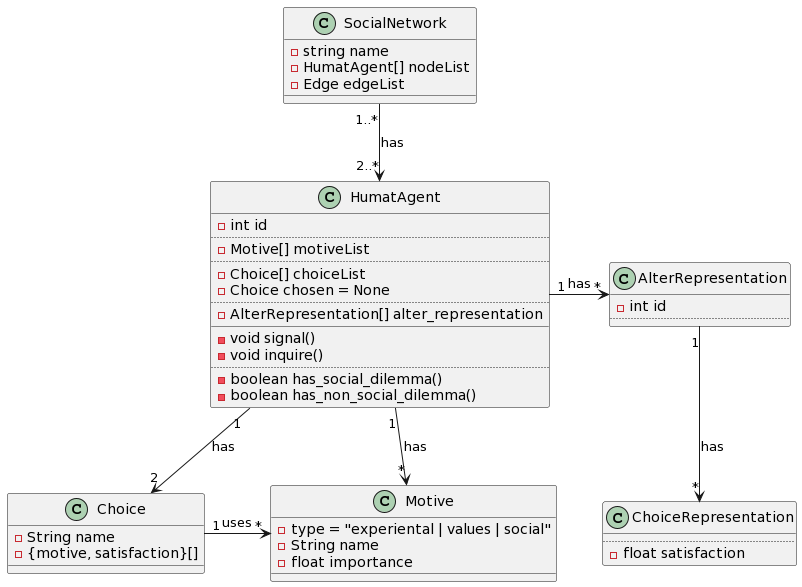}
\caption{The generic conceptual UML model for HUMAT.}
\label{fig:HUMAT-Domain-Model}
\end{figure}

To do this effectively, each HUMAT has a representation of all alters linked to it in the ego network. 
An activated link between HUMAT and the targeted alter denotes a communication act - sharing information about the subject of the attitude (either inquiring or signaling). 
The persuasiveness of the communicating agent depends on similarity and aspirational characteristics relevant to a given research context. 
%For a detailed description of the HUMAT architecture see \cite{antosz_simulation_2019}, \cite{antosz_smartees_2021}, \cite{czupryna_documenting_2022}, \cite{squazzoni_hum-e_nodate}.

\begin{figure}
\centering
\includegraphics[width=0.99\textwidth]{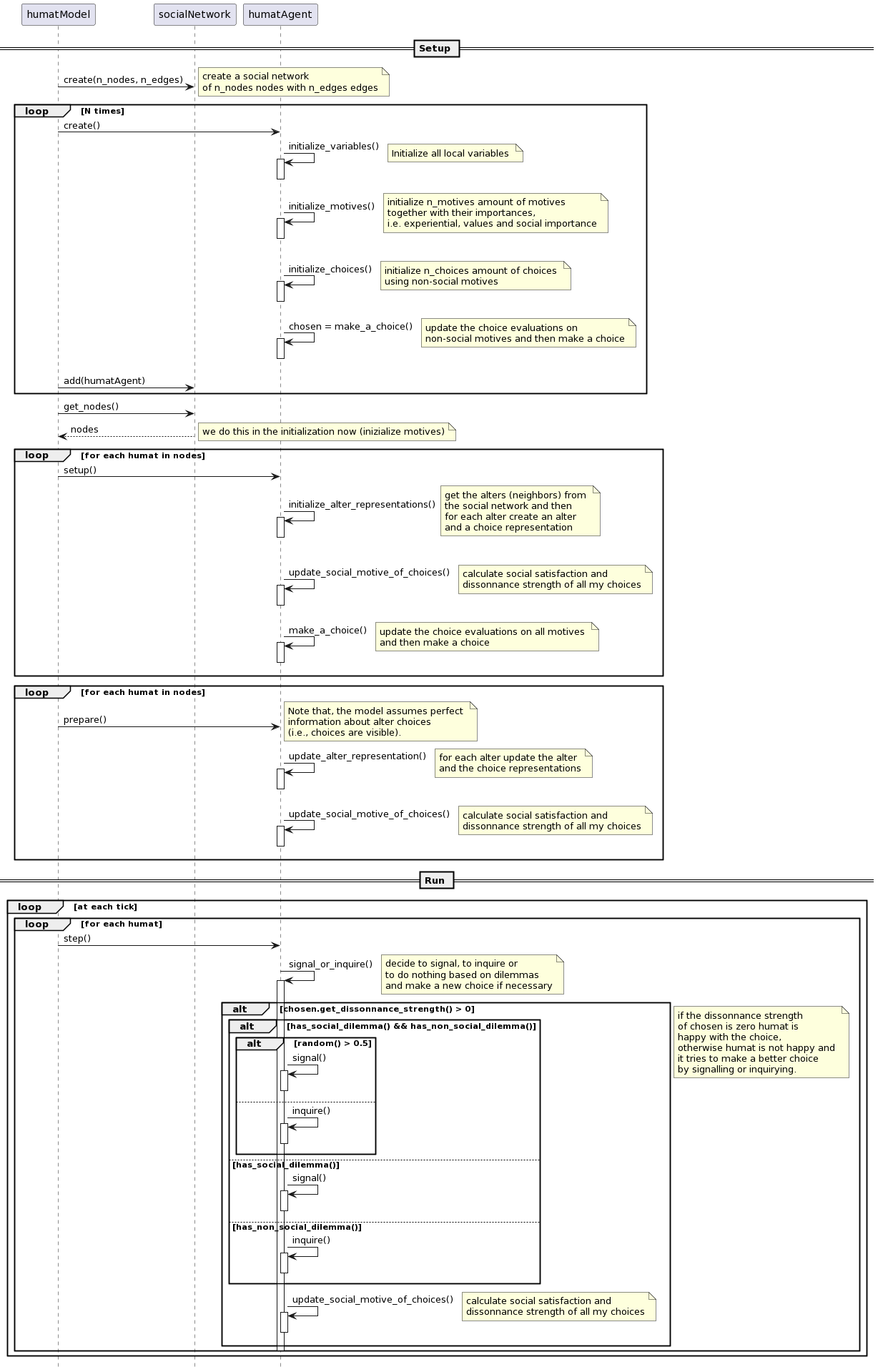}
\caption{The generic behavioral UML model for HUMAT.}
\label{fig:HUMAT-Behavioral-Model}
\end{figure}

Based on the above understanding, we designed a programming language-independent generic model for HUMAT (Figure \ref{fig:HUMAT-Domain-Model} and Figure \ref{fig:HUMAT-Behavioral-Model}). 
Figure \ref{fig:HUMAT-Domain-Model} depicts the high-level representations of various concepts in the HUMAT domain and their relationships.
Figure \ref{fig:HUMAT-Behavioral-Model} represents an overall behavioral model for a HUMAT agent within a social network. The model initializes nodes (HUMATS) and edges in the social network, creating agent instances, and initializing their variables, motives, and choices. Then, it adds the agents to the network, initializes their representations of other agents (alters), and updates their social motives for choices.
During each simulation step (tick), agents may decide to signal, inquire, or do nothing based on their current dilemmas and the dissonance strength of their chosen action. If an agent is not satisfied with their choice, they will try to become more content by signalling or inquiring. The basic version of the HUMAT architecture assumes perfect information about alter choices, meaning that all choices are visible to other agents in the ego network.
Throughout the simulation, the agents continuously update their alter representations, social motives of choices and make new choices based on their evaluations of motives and dissonance strength.

\subsection{Reimplementing HUMAT in Python}

\begin{figure}
\centering
\includegraphics[width=0.99\textwidth]{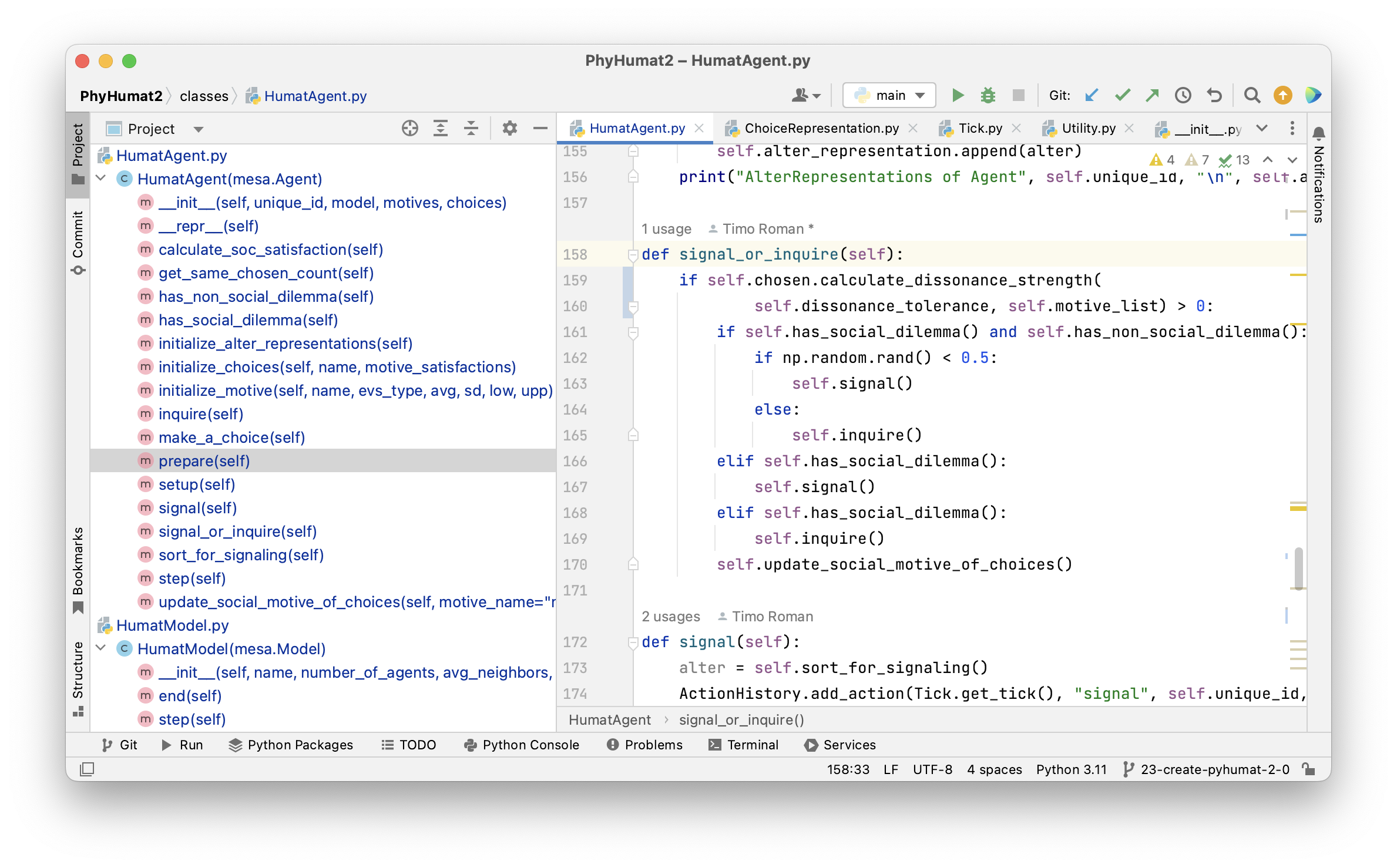}
\caption{Python code for the Signal or Inquire function.}
\label{fig:SignalOrInquire}
\end{figure}

Re-implementing HUMAT in Python from the generic conceptual model is a straightforward process. 
Each concept in the generic model is translated into a Python class with related parameters and methods. 
The two main classes of the model describe the agents (HumatAgent) and the model (HumatModel) (see Figure \ref{fig:SignalOrInquire}). 

The HumatModel class extends the Mesa Model class and controls methods executed during a time step.
The HumatAgent class extends the Mesa Agent class and controls the methods executed by the agent. 
%One of the features of the Mesa package is the Agent and the Model classes. 
The generic model does not specify which Python data types, syntax, and packages to use. 
These decisions are up to the modeler and depend on their personal experiences.

\subsection{Validating HUMAT in Python}
% Validation procedures play a crucial role in model re-implementation. While the specifics of these procedures may vary, in our case they aim to achieve numerical identity. 
We start each validation by configuring both models identically by importing all model states of the NetLogo model into Python after initial initialization. 
Subsequently, automatic unit tests of agent parameters are executed at each time step. 
This process is repeated, considering increasing agent populations and degrees of randomization (e.g., by controlling scheduling). 
Throughout the testing, the methods' functionality is reviewed and optimized. The findings of this comprehensive case study will be documented in a separate paper.

\section{Discussion}

Replicating a NetLogo model directly in Python poses some specific challenges in the implementation process. A brief description of the main challenges we faced is given below. More complete comparisons between NetLogo and Python can be found in \cite{abbott_pylogo_2021}.

\begin{itemize}
    \item  Object-oriented coding and methods: The NetLogo model is written as a collection of procedures: (i) a setup defines agent parameters and the model environment (e.g., patches and networks), (ii) the main go loop is then executed to run all model procedures for a defined number of time steps (ticks), (iii) the remaining procedures are non-restrictive and written anywhere below the setup and go procedure. 
    The Python model is organized into classes with specific methods: (i) the main class contains methods defining the model inputs and the number of time steps, (ii) the model class controls methods executed during a time step, (iii) the §agent class control the methods that agents execute.
    In NetLogo, two types of procedures are: \textit{to} and \textit{to-report}. The \textit{to} procedures usually contain a set of commands executed (e.g., by agents), while the \textit{to-report} returns a value. Functions and methods in Python can execute commands or return values. 
    \item Turtles and Breed vs. Agent classes: NetLogo has four predefined agent types: turtles, patches, links, and the observer. Breeds are used to define specific sub-groups of agents (e.g., HUMATS). Each agent and breed can have specific parameters assigned to it and can be controlled using NetLogo keywords, the primitives. 
    Python, on the other hand, uses classes to define objects. One of the features of the Mesa is the Agent class. Each object created as a sub-class of Agent is automatically equipped with a unique \textit{id} and a \textit{step()} method and inherits features.
    \item Agentset, lists, and dictionaries: In NetLogo, groups of agents are organised in agentsets. These sets of agents can be created on the fly in random order. Agentsets are a very comfortable way to control or select a subset of agents using a set of primitives. While Python can create agent sets, storing agents in dictionaries is often more convenient. 
    
\end{itemize}

Due to these challenges, it is not practical to re-implement a model in Python directly from a NetLogo model. 
The difference in abstractions used in both languages will make it hard for the modeler to transition. 
Consequently, for an effective re-implementation and rapid re-implementation in other programming languages, abstracting away the programming language concept and designing a generic model is essential.
For instance, thanks to this generic model, we plan to re-implement HUMAT in Java for another project, and it will be pretty rapid.

\section{Conclusions and Future Work}

This paper contributes to the literature in three meaningful ways.
One, previous studies agree on the importance of replicating agent-based models, however they mostly present experiences on individual models (e.g., \cite{an_agent-based_2009,thiele_facilitating_2014,liang_testing_2015}). 
Here, we add to the existing general guidelines \cite{zhang_replication_2021} by proposing a programming language-independent systematic approach, from understanding the existing model to sharing the new implementation. 

Two, replications of ABMs focus on discussing the validation of the re-implemented model: to what extent the outputs of the re-implemented model are aligned with the outputs of the original model \cite{axelrod_advancing_1997}. 
The case study of replicating HUMAT described here focuses on the re-implementation process rather than the model outcomes.

Three, the authors provide a glimpse of the re-implementation process report having developed a general, conceptual model that is the basis of the original ABM \cite{wilensky_making_2007} or a platform-independent model \cite{sansores_agent-based_2005}. This is a similar approach to developing a generic model proposed here. 
An intermediate, generic model enables a focus on the investigated phenomenon without anchoring in the concepts present in a given programming language. Additionally, it makes further re-implementations in different languages faster and less effortful.

Up until now, we closely followed the guideline until the \textit{Validate the New Model} step. 
This remaining step involves sensitivity analysis and testing of the new model and thus requires a more detailed discussion.
In future work, we will finish validating the new model implemented in Python and report the results of our experience. 
%This is an experience paper, so we can speak about the difficulties we encountered, and speaking about following a method is essential and the literature lacks such methods and so on. If there are some methods already, we need to mention them and compare them with ours shortly.

We hope that, in future, the guidelines will be used and perfected by the social simulation community. 
To make re-implementation of ABMs more common, modellers should follow the \textit{Share and collaborate} step of the proposed guideline. The social simulation community can popularize such works by initiating a dedicated label in COMSES or launching a publication outlet focusing on model evaluation, replication and re-implementation.

\subsubsection{Acknowledgements} The work reported here is part of the URBANE project, which has received funding from the European Union’s Horizon Europe Innovation Action under grant agreement No. 101069782. We thank the reviewers for the thoughtful remarks, especially related to the popularization ideas.

\bibliographystyle{splncs04}
\bibliography{references}

\begin{thebibliography}{10}
\providecommand{\url}[1]{\texttt{#1}}
\providecommand{\urlprefix}{URL }
\providecommand{\doi}[1]{https://doi.org/#1}

\bibitem{abar_agent_2017}
Abar, S., Theodoropoulos, G.K., Lemarinier, P., O’Hare, G.M.P.: Agent {Based}
  {Modelling} and {Simulation} tools: {A} review of the state-of-art software.
  Computer Science Review  \textbf{24},  13--33 (2017)

\bibitem{abbott_pylogo_2021}
Abbott, R., Lim, J.: {PyLogo}: {A} {Python} {Reimplementation} of ({Much} of)
  {NetLogo}:. In: Proceedings of the 11th {International} {Conference} on
  {Simulation} and {Modeling} {Methodologies}, {Technologies} and
  {Applications}. pp. 199--206. SCITEPRESS - Science and Technology
  Publications, Online Streaming (2021)

\bibitem{achter_rat-rs_2022}
Achter, S., Borit, M., Chattoe-Brown, E., Siebers, P.O.: {RAT}-{RS}: a
  reporting standard for improving the documentation of data use in agent-based
  modelling. International Journal of Social Research Methodology
  \textbf{25}(4),  517--540 (Jul 2022)

\bibitem{an_agent-based_2009}
An, G., Mi, Q., Dutta-Moscato, J., Vodovotz, Y.: Agent-based models in
  translational systems biology. Wiley interdisciplinary reviews. Systems
  biology and medicine  \textbf{1}(2),  159--171 (2009)

\bibitem{antosz_simulation_2019}
Antosz, P., Jager, W., Polhill, J.G., Salt, D., Alonso-Betanzos, A.,
  Sánchez-Maroño, N., Guijarro-Berdiñas, B., Rodríguez, A.: Simulation
  model implementing different relevant layers of social innovation, human
  choice behaviour and habitual structures. Tech. Rep.~D7.2 (2019)

\bibitem{antosz_smartees_2021}
Antosz, P., Jager, W., Polhill, J.G., Salt, D., Alonso-Betanzos, A.,
  Sánchez-Maroño, N., Guijarro-Berdiñas, B., Rodríguez, A., Scalco, A.:
  {SMARTEES} simulation implementations. Tech. Rep.~D7.3 (2021)

\bibitem{czupryna_documenting_2022}
Antosz, P., Puga-Gonzalez, I., Shults, F.L., Lane, J.E., Normann, R.:
  Documenting {Data} {Use} in a {Model} of {Pandemic} “{Emotional}
  {Contagion}” {Using} the {Rigour} and {Transparency} {Reporting} {Standard}
  ({RAT}-{RS}). In: Czupryna, M., Kamiński, B. (eds.) Advances in {Social}
  {Simulation}, pp. 439--451. Springer, Cham (2022)

\bibitem{squazzoni_hum-e_nodate}
Antosz, P., Puga-Gonzalez, I., Shults, F.L., Szczepanska, T.: {HUM}-e: {An}
  emotive-socio-cognitive agent architecture for representing human
  decision-making in anxiogenic contexts. In: Squazzoni, F. (ed.) Advances in
  {Social} {Simulation}. Springer International Publishing, Cham

\bibitem{antosz_sensemaking_2022}
Antosz, P., Szczepanska, T., Bouman, L., Polhill, J.G., Jager, W.: Sensemaking
  of causality in agent-based models. International Journal of Social Research
  Methodology  \textbf{25}(4),  557--567 (Jul 2022).
  \doi{10.1080/13645579.2022.2049510}

\bibitem{axelrod_advancing_1997}
Axelrod, R.: Advancing the {Art} of {Simulation} in the {Social} {Sciences}.
  In: Simulating {Social} {Phenomena}, Lecture {Notes} in {Economics} and
  {Mathematical} {Systems}, vol.~456, pp. 21--40. Springer Berlin Heidelberg,
  Berlin, Heidelberg (1997)

\bibitem{de_bok_empirical_2018}
de~Bok, M., Tavasszy, L.: An empirical agent-based simulation system for urban
  goods transport ({MASS}-{GT}). Procedia Computer Science  \textbf{130},
  126--133 (2018)

\bibitem{chattoe-brown_reproduction_2021}
Chattoe-Brown, E., Gilbert, N., Robertson, D.A., Watts, C.: Reproduction as a
  {Means} of {Evaluating} {Policy} {Models}: {A} {Case} {Study} of a {COVID}-19
  {Simulation}. medRxiv  (2021). \doi{10.1101/2021.01.29.21250743}

\bibitem{collier_repast_2003}
Collier, N.: {RePast}: {An} {Extensible} {Framework} for {Agent} {Simulation}.
  The University of Chicago’s Social Science Research  (2003)

\bibitem{edmonds_replication_2003}
Edmonds, B., Hales, D.: Replication, {Replication} and {Replication}: {Some}
  {Hard} {Lessons} from {Model} {Alignment} (Oct 2003)

\bibitem{grimm_pattern-oriented_2012}
Grimm, V., Railsback, S.F.: Pattern-oriented modelling: a ‘multi-scope’ for
  predictive systems ecology. Philosophical Transactions of the Royal Society
  B: Biological Sciences  \textbf{367}(1586),  298--310 (2012)

\bibitem{grimm_odd_2020}
Grimm, V., Railsback, S.F., Vincenot, C.E., Berger, U., Gallagher, C.,
  DeAngelis, D.L., Edmonds, B., Ge, J., Giske, J., Groeneveld, J., Johnston,
  A.S.A., Milles, A., Nabe-Nielsen, J., Polhill, J.G., Radchuk, V., Rohwäder,
  M.S., Stillman, R.A., Thiele, J.C., Ayllón, D.: The {ODD} {Protocol} for
  {Describing} {Agent}-{Based} and {Other} {Simulation} {Models}: {A} {Second}
  {Update} to {Improve} {Clarity}, {Replication}, and {Structural} {Realism}.
  Journal of Artificial Societies and Social Simulation  \textbf{23}(2), ~7
  (2020)

\bibitem{gurcan_towards_2011}
G\"{u}rcan, O., Dikenelli, O., Bernon, C.: Towards a {Generic} {Testing}
  {Framework} for {Agent}-{Based} {Simulation} {Models}. In: Ganzha, M.,
  Maciaszek, L.A., Paprzycki, M. (eds.) {FedCSIS} 2011. pp. 635--642. Szczecin,
  Poland (Sep 2011)

\bibitem{gurcan_generic_2014}
G\"{u}rcan, O., Dikenelli, O., Bernon, C.: A generic testing framework for
  agent-based simulation models. In: Agent-{Based} {Modeling} and {Simulation},
  pp. 231--270. Springer (2014)

\bibitem{larman_applying_2004}
Larman, C.: Applying {UML} and {Patterns}: {An} {Introduction} to
  {Object}-{Oriented} {Analysis} and {Design} and {Iterative} {Development}
  (3rd {Edition}). Prentice Hall, USA (2004)

\bibitem{liang_testing_2015}
Liang, H., Fu, K.w.: Testing {Propositions} {Derived} from {Twitter} {Studies}:
  {Generalization} and {Replication} in {Computational} {Social} {Science}.
  PLOS ONE  \textbf{10}(8),  1--14 (Aug 2015).
  \doi{10.1371/journal.pone.0134270}

\bibitem{masad_mesa_2015}
Masad, D., Kazil, J.: Mesa: {An} {Agent}-{Based} {Modeling} {Framework}. pp.
  51--58. Austin, Texas (2015). \doi{10.25080/Majora-7b98e3ed-009}

\bibitem{maxwell_is_2015}
Maxwell, S.E., Lau, M.Y., Howard, G.S.: Is psychology suffering from a
  replication crisis? Am Psychol.  \textbf{70}(6),  487--498 (2015)

\bibitem{north_agent_2009}
North, M.J., Macal, C.M.: Agent {Based} {Modeling} and {Computer} {Languages}.
  In: Meyers, R.A. (ed.) Encyclopedia of {Complexity} and {Systems} {Science},
  pp. 131--148. Springer New York (2009). \doi{10.1007/978-0-387-30440-3_8}

\bibitem{pressman_software_2014}
Pressman, R., Maxim, B.: Software {Engineering}: {A} {Practitioner}’s
  {Approach}, 8th {Ed} (Jan 2014)

\bibitem{railsback_agent-based_2019}
Railsback, S., Grimm, V.: Agent-{Based} and {Individual}-{Based} {Modeling}:
  {A} {Practical} {Introduction}. Agent-based and {Individual}-based
  {Modeling}: {A} {Practical} {Introduction}, Princeton University Press (2019)

\bibitem{railsback_concepts_2001}
Railsback, S.F.: Concepts from complex adaptive systems as a framework for
  individual-based modelling. Ecological Modelling  \textbf{139}(1),  47--62
  (2001)

\bibitem{sansores_agent-based_2005}
Sansores, C., Pavón, J.: Agent-{Based} {Simulation} {Replication}: {A} {Model}
  {Driven} {Architecture} {Approach}. In: {MICAI} 2005: {Advances} in {AI} 4th
  {Mexican} {Int}. {Conf}. on {AI}, {LNAI}, vol.~3789, pp. 244--253. Springer
  (2005)

\bibitem{tang_code_2020}
Tang, W., Grimm, V., Tesfatsion, L., Shook, E., Bennett, D., An, L., Gong, Z.,
  Ye, X.: Code {Reusability} and {Transparency} of {Agent}-{Based} {Modeling}:
  {A} {Review} from a {Cyberinfrastructure} {Perspective}. In: Tang, W., Wang,
  S. (eds.) High {Performance} {Computing} for {Geospatial} {Applications}, pp.
  115--134. Springer, Cham (2020)

\bibitem{thiele_facilitating_2014}
Thiele, J.C., Kurth, W., Grimm, V.: Facilitating {Parameter} {Estimation} and
  {Sensitivity} {Analysis} of {Agent}-{Based} {Models}: {A} {Cookbook} {Using}
  {NetLogo} and '{R}'. Journal of Artificial Societies and Social Simulation
  \textbf{17}(3), ~11 (2014)

\bibitem{wilensky_making_2007}
Wilensky, U., Rand, W.: Making {Models} {Match}: {Replicating} an
  {Agent}-{Based} {Model}. Journal of Artificial Societies and Social
  Simulation  \textbf{10}(4) (2007)

\bibitem{will_replication_2008}
Will, O., Hegselmann, R.: A {Replication} {That} {Failed}: {On} the
  {Computational} {Model} in '{Michael} {W}. {Macy} and {Yoshimichi} {Sato}:
  {Trust}, {Cooperation} and {Market} {Formation} in the {U}.{S}. and {Japan}.
  JASSS  \textbf{11}(3) (2008)

\bibitem{zhang_replication_2021}
Zhang, J., Robinson, D.T.: Replication of an agent-based model using the
  {Replication} {Standard}. Environmental Modelling \& Software  \textbf{139},
  105016 (2021)

\bibitem{zhong_using_2010}
Zhong, W., Kim, Y.: Using {Model} {Replication} to {Improve} {Reliability} of
  {Agent}-{Based} {Models}. In: Chai, S.K., Salerno, J.J., Mabry, P.L.,
  Hutchison, D., Kanade, T. (eds.) Advances in {Social} {Computing}: 3rd {Int}.
  {Conf}. on {Social} {Computing}, {Behavioral} {Modeling}, and {Prediction},
  {LNCS}, vol.~6007, pp. 118--127. Springer (2010)

\end{thebibliography}
\end{document}